\documentclass[11pt]{article}
\usepackage{latexsym}  % for Symbol fonts
\usepackage{amssymb}
\usepackage{graphicx}
\usepackage{amsmath}
\begin{document}

\begin{center}

{\Large\bf Sumino Model and My Personal View}\footnote{
A talk given at a mini-workshop on ``quarks, leptons and family gauge 
bosons", Osaka University, Osaka, Japan, December 26-27, 2016.
}.

\vspace{4mm}

{\bf Yoshio Koide}

{\it Department of Physics, Osaka University, 
Toyonaka, Osaka 560-0043, Japan} \\
{\it E-mail address: koide@kuno-g.phys.sci.osaka-u.ac.jp}

%\vspace{2mm}
\date{\today}
\end{center}

\vspace{3mm}
%\maketitle

\begin{abstract}
There are two formulas for charged lepton mass relation: 
One is a formula (formula A) which was proposed based on 
a U(3) family model on 1982.   The formula A will be satisfied 
only masses switched off all interactions except for U(3) 
family interactions.  Other one (formula B) is an empirical 
formula which we have recognized after a report of the precise 
measurement of tau lepton mass, 1992.  The formula B is 
excellently satisfied by pole masses of the charged leptons.  
However, this excellent agreement may be an accidental 
coincidence.  Nevertheless, 2009, Sumino has paid attention 
to the formula B.  He has proposed a family gauge boson model 
and thereby he has tried to understand why the formula B 
is so well satisfied with pole masses.  In this talk, 
the following views are given: (i) What direction 
of flavor physics research is suggested by the formula A; 
(ii) How the Sumino model is misunderstood by people and 
what we should learn from his model; (iii) What is strategy 
of my recent work, U(3)$\times$U(3)$'$ model.
\end{abstract}

%\pacs{
%PCAC numbers:  
%  11.30.Hv, % Flavor symmetries 
%  12.15.Ff, %Quark and lepton masses and mixing
%  14.60.Pq,  % Neutrino mass and mixing
% 12.60.-i, % Models beyond the standard model
%}
%%%%%%%%%%%%%%%%%

\vspace{5mm}

\noindent{\bf 1  \  Two formulas for charged lepton masses} 

Prior to discussing the Sumino model \cite{Sumino_PLB09}, 
let us review a charged lepton mass relation, 
We know two formulas for the charged lepton masses.  
One is a formula (formula A) which was proposed based on 
a U(3) family model on 1982 \cite{K-mass_82}: 
$$
K(m_{ei} ) \equiv \frac{m_e + m_\mu +m_\tau}{\left(\sqrt{m_e}+ 
\sqrt{m_\tau} +\sqrt{m_\tau}\right)^2} = \frac{2}{3} . 
\eqno(1)
$$  
The formula A will be satisfied only masses which are given 
in the world switched off all interactions except for the U(3) 
family interactions.  
Other one (formula B) is an empirical 
formula which we have recognized since precise observation 
of tau lepton mass \cite{tau-mass_92}, 1992:   
$$ 
K(m_{ei}^{pole})  = \frac{2}{3} 
\times(0.999989 \pm 0.000014) .
\eqno(2)
$$
The formula B is excellently satisfied with pole masses 
of the charged leptons.
However, this excellent agreement may be an accidental 
coincidence.  

I regret that some people simple-honestly 
tried to search mathematical quantities which leads 
to the form $K=2/3$.  
They did not understand that $m_{ei}$ in the 
formula B are pole masses, and besides, that the mass 
spectra cannot discuss independently of the flavor mixing.    
Most of their attempts could not left any physical 
result.
Their attempts are nothing but playing of a mathematical 
puzzle, not physics. 

Independently of whether the formula A can give well numerical 
agreement or not, if we accept the formula A, we will also 
accept the following points of view:

\noindent (i) We know the quark mixing and neutrino mixing.  
Therefore, the formula A holds only in the charged lepton
sector, so that a similar relation never hold in other sectors 
(up-quark, down-quark and neutrino sectors).
In other words, we should discuss flavor physics 
on the diagonal bases of the charged lepton mass 
matrix $M_{e}$. 

\noindent (ii) Masses and mixings should be investigated   
based on $M_e^{1/2}$, not $M_e$. 

\noindent \noindent (iii) The observed hierarchical mass 
spectra in quarks and leptons suggest that those cannot be 
understood from a conventional symmetry approach 
(symmetry + a small breaking), and 
it has to be understood form vacuum expectation 
values (VEVs) of scalars with Higgs-like mechanism
\cite{YK_MPLA90, Yukawaon}.

\noindent (iv) If we put $m_e=0$ in the formula A, we obtain 
unwelcome ratio of $m_\mu/m_\tau=1/(2+\sqrt{3})$. 
This suggests that we have to seek for mass matrix 
model in which the electron mass should be given by a 
non-zero value from the beginning even if it is very small.  
The mass spectrum $(m_e, m_\mu, m_\tau)$ has to be 
understood simultaneously, that is, without considering 
a mass generation model with two or three steps.

%%%%%%%%%%%%%%%%%%%%%%%%%%%%%%%%%%%%%%%%%%%%%%%%%%%%%%%%%%%%%%%%
\vspace{5mm}

\noindent{\bf 2  \  Impact of the Sumino model} 

However, against such my personal view given in the previous 
section, 2009, Sumino \cite{Sumino_PLB09} has paid 
attention to the formula B.  
He has proposed a family gauge boson model and thereby, 
he has tried to understand why the formula B is so well 
satisfied with pole masses. 

The formula A is invariant under a transformation
$$
m_{ei} \ \rightarrow \ m_{ei}(1 + \varepsilon_0 ) ,
\eqno(3) 
$$
where $\varepsilon_0$ is a constant which is independent
of the family-number.
The deviation from the formula A due to QED correction 
comes from $\log m_{ei}$:
$$
\delta m_{ei} = m_{ei} \left( 1 + c_1^{QED} \log m_{ei} +
c_0^{QED} \right) ,
\eqno(4)
$$  
at the level of the one-loop correction \cite{Arason_PRD92}. 
Therefore, Sumino has assumed an existence of family gauge 
bosons (FGBs) $A_i^{ j}$ with the masses 
$M_{ij}^2 =k (m_{ei} + m_{ej})$ 
and thereby, he has proposed a cancellation mechanism 
between $\log m_{ei} $ in the QED correction and $\log M_{ii}$ 
in the FGB one-loop correction:
$$
\delta m_{ei} = m_{e_i} \left( c_1^{QED} \log m_{ei} + 
c_1^{FGB} \log M_{ii} + const \right) = m_{ei} \times const.
\eqno(5)
$$
 
His model is based on U(3)$\times$O(3) symmetry. 
In his model, in order to obtain a minus sign for the 
cancellation, the quarks and leptons $f$ are assigned 
to $(f_L, f_R)=({\bf 3}, {\bf 3}^*)$ of U(3), so that  
the model is not anomaly free.
Besides, effective interactions with $\Delta N_{fam} = 2$
 ($N_{fam}$ is a family number) appears.
Therefore, in order to remove those shortcomings, an 
extended Sumino model has been proposed based on 
U(3)$\times$U(3)$'$ symmetry
and with an inverted mass hierarchy of FGBs \cite{K-Y_PLB12}. 
However, the purpose of this talk is not to review those 
details.

The big objection is that there are many diagrams which 
we should take into consideration, so that the Sumino 
cancellation mechanism cannot work effectively.    
However, it is misunderstanding for the Sumino mechanism.  
Sumino has already taken those effects into 
consideration. 
The Sumino cancellation mechanism does not mean complete   
cancellation, but it means practical cancellation 
at a level of the present experimental accuracy. 
In fact, Sumino says that if accuracy in the 
present tau lepton mass measurement can be improved 
to one order, the deviation from the formula B will 
be observed. 
Also, he has said that the upper bound in which the 
cancellation mechanism is effective is $10^3$-$10^4$ TeV. 
We hope that the soon coming tau mass measurement will 
verify Sumino's conjecture. 

In his model, the masses $M_{ij}$ are related to 
the charged lepton masses $m_{ei}$, and the family 
gauge coupling constant $g_F$ is related to 
QED gauge coupling constant $e$. 
Therefore, the FGB model has highly predictability.  

The most notable point of Sumino FGB model is that there 
is a upper limit of the FGB mass scale, which comes 
from applicability of the Sumino mechanism. 
Therefore, when Sumino FGBs cannot be discovered 
at the expected scale, we cannot excuse the undiscovered 
fact by extending the scale to one order. 
In such a case, we have to abandon the Sumino model.

Even apart from the Sumino cancellation mechanism, 
his FGB model has many notable characteristics. 
In his model, the FGB masses $M_{ij}$ and the 
charged lepton masses $m_{ei}$ are generated by 
the same scalar $\Phi =({\bf 3}, {\bf 3})$
of U(3)$\times$O(3), so that, when the charged lepton 
mass matrix $M_e$ is diagonal, the FGB mass matrix is 
also diagonal.
Therefore, family-number violation does not occur in
the lepton sector.
Family-number violation appears only in the quark sector
only via quark mixing. 
Therefore, in the limit of zero quark mixing, 
family-number violation disappears in the quark sector, too.
Thus, the Sumino family FGB model offers us FGBs with
considerably low scale without constraining the 
conventional view from the observed $K^0$-$\bar{K}^0$ 
mixing and so on \cite{YK_PLB14}.  
Now we may expect observations of FGBs in terrestrial 
experiments.  
We will have fruitful new physics related to Sumino 
FGB model.

%%%%%%%%%%%%%%%%%%%%%%%%%%%%%%%%%%%%%%%%%%%%%%%%%%%%%%%%%
\vspace{5mm}

\noindent{\bf 3  \  Strategy of the U(3)$\times$U(3)$'$ model} 

 Stimulating by the Sumino model, I have recently investigated 
a unified description of quarks and leptons based on 
U(3)$\times$U(3)$'$ symmetry \cite{K-N_PRD15R, K-N_MPLA16}.  
Here, quarks and leptons are assigned to $({\bf 3},{\bf 1})$ 
of U(3)$\times$U(3)$'$. 
Nevertheless, we need additional symmetry U(3)$'$ with 
considerably higher scale. 
Why?  The reason will see an example in the following mass 
matrix model (although our interest is not only in masses 
and mixing). 

In order to give a review of the U(3)$\times$U(3)$'$ model 
concretely, let us take a mass matrix model based on 
the U(3)$\times$U(3)$'$ symmetry.
In this model, heavy fermions $F_\alpha$ ($\alpha=1,2,3$) 
are introduced in addition to quarks and leptons $f_i$ 
($i=1,2,3$). 
$F_\alpha$ and $f_i$ belong to $({\bf 1}, {\bf 3})$ and 
$( {\bf 3}, {\bf 1})$ of U(3)$\times$U(3)$'$, respectively. 
We consider a seesaw-like mass matrix:
$$
(\bar{f}_L^i \ \ \bar{F}_L^\alpha ) 
\left(
\begin{array}{cc}
(0)_i^{\ j}  &  (\Phi_f)_i^{\ \beta}  \\
(\bar{\Phi}_{f})_\alpha^{\ j} & -(S_f)_\alpha^{\ \beta} 
\end{array} \right) 
 \left(
\begin{array}{c}
f_{Rj} \\
F_{R\beta}
\end{array} 
\right) .
\eqno(6)
$$
Since we consider that U(3)$'$ is broken into a discrete 
symmetry S$_3$, a VEV form of $\hat{S}_f$, in general, 
takes a form (unit matrix + democratic matrix):
$$
\langle \hat{S}_f \rangle = v_{S} ({\bf 1} + b_f X_3) ,
\eqno(7)
$$
where ${\bf 1}$ and $X_3$ are defined as
$$
{\bf 1} = \left( 
\begin{array}{ccc}
1 & 0 & 0 \\
0 & 1 & 0 \\
0 & 0 & 1 
\end{array} \right) , \ \ \ \ \ 
X_3 = \frac{1}{3} \left( 
\begin{array}{ccc}
1 & 1 & 1 \\
1 & 1 & 1 \\
1 & 1 & 1 
\end{array} \right) ,  
\eqno(8)
$$
and $b_f$ are complex parameters.  
On the other hand, we consider that U(3) is broken by 
VEVs of $\Phi_f$ with $({\bf 3}, {\bf 3}^*)$ of 
U(3)$\times$U(3)$'$, not by $({\bf R}, {\bf 1})$
of U(3)$\times$U(3)$'$.
Here, the VEV forms are diagonal and given by
$$
\langle \Phi_f \rangle = v_\Phi\, {\rm diag}( z_1 e^{i\phi^f_1}, 
z_2 e^{i\phi^f_2}, z_3 e^{i\phi^f_3}) .
\eqno(9)
$$
Since we consider $|\langle\hat{S}_f\rangle | \gg |\langle \Phi_f 
\rangle |$, we obtain a seesaw-like Dirac mass matrix for $f$ 
\cite{K-F_ZPC96}
$$
(\hat{M}_f)_i^{\ j}  = \langle\Phi_f\rangle_i^{\ \alpha}
\langle \hat{S}_f^{-1} \rangle_\alpha^{\ \beta} 
\langle \bar{\Phi}_f\rangle_\beta^{\ j} .
\eqno(10)
$$
Since our model gives $b_e=0$ for the charged lepton sector,
so that the charged lepton mass matrix is diagonal,  
the parameters $z_i$ given in Eq.(9) can be expressed as 
$$
z_i = \frac{ \sqrt{m_{ei}} }{\ \sqrt{ 
m_e + m_\mu + m_\tau} } .
\eqno(11)
$$
As a result, masses and mixings of quarks and neutrinos 
are only the family-number independent parameters $b_f$.    
(We will take the phase factors $\phi^f_i$ as $\phi^f_i=0$ 
except for $f=u$.)  
Even for the family-number dependent parameters $\phi^u_i$, 
we can express those by 
the parameters $(z_1,z_2, z_3)$ and two family-number  
independent parameters \cite{K-N_PRD15}.)  
Thus, masses and mixings of quarks and leptons are 
governed by rules in U(3)$\times$U(3)$'$, not
in U(3).  

Note that we have used the observed values of charged 
lepton masses for the parameters $z_i$ given in Eq.(11). 
We never ask any origin of the charged lepton mass 
spectrum $(m_e, m_\mu, m_\tau)$. 
Our strategy is as follows: our aim is to describe
quarks and neutrino masses and mixings only by using 
the observed values $(m_e, m_\mu, m_\tau)$ and 
without using any family-number dependent parameters.  
We consider that it is too early to investigate the 
origin of $(m_e, m_\mu, m_\tau)$,
that is, U(3) symmetry breaking mechanism.  
It is a future task to us. 
 
However, there are still many remaining tasks in the
U(3)$\times$U(3)$'$ model. 
We have to improved this model into more simple and 
reliable model.  
(For a recent work in the U(3)$\times$U(3)$'$ model, 
for example, see Ref.\cite{K-N_17}.)

\vspace{5mm}
%

%%%%%%%%%%%%%%%%%%%%%%%%%%%%%%%

%    


\begin{thebibliography}{99} %%%%%%%%%%%%%%%%%
%%%%%%%%%%%%%%%%%%%%%%%%%%%%%%%
%
%
\bibitem{Sumino_PLB09}
Y.~Sumino, Phys. Lett. {\bf B 671} (2009) 477; JHEP 0905 (2009).
%
\bibitem{K-mass_82} 
  Y.~Koide,
  Lett.\ Nuovo Cim.\  {\bf 34} (1982)  201;
  Phys.\ Lett.\  B {\bf 120}(1983)  161;
  Phys.\ Rev.\  D {\bf 28} (1983) 252.
%  For a recent work, for see, 
%Y.~Koide, Phys.\ Rev.\  D {\bf 79} (2009) 033009.
% 
\bibitem{tau-mass_92}
H.~Albrecht {\it et al.} ARGUS collab., Phys.=Lett. {\bf B292}, 
221 (1992); 
J.~Z.~Bai {\it el al.} BES collab., Phys.~Rev.~Lett. {\bf 69}, 3021
(1982); 
M.~Daoudi {\it et al.} CLEO collab., a talk given at the XXVI 
int.~Conf.~on High Energy Physics, Dallas, 1992. 
%
%
\bibitem{YK_MPLA90}
Y.~Koide, Mod.~Phys.~Lett. A {\bf 28}, 2319 (1990). 
%
\bibitem{Yukawaon}
Y.~Koide, Phys.~Rev. {\bf D 79} (2009), 033009; Phys.~Lett. {\bf B 680}
(2009) 76. 
%

\bibitem{Arason_PRD92}
H.~Arason, {\it et al}., Phys.~Rev. {\bf D 46}, 3945 (1992).
%% 
%
\bibitem{K-Y_PLB12}
Y.~Koide and T.~Yamashita, 
% ``Family gauge bosons with an inverted mass hierarchy",
  Phys.\ Lett.\  B {\bf 711}, 384 (2012).
  %%CITATION = PHLTA,B711,384;%%
%  [arXiv:1203.2028 [hep-ph].
%
%
\bibitem{YK_PLB14} 
Y.~Koide,  Phys.\ Lett.\  B {\bf 736}, 499 (2014).
%
%
\bibitem{K-N_PRD15R}
Y.~Koide and H.~Nishiura,  Phys.Rev. {\bf D 92} (2015) 111301(R).
%
\bibitem{K-N_MPLA16}
Y.~Koide and H.~Nishiura, Mod.Phys.Lett. A {\bf 31} (2016) 1650125. 
%
%
\bibitem{K-F_ZPC96} 
Y.~Koide and H.~Fusaoka, Z.~Phys. {\bf C 71} (1996) 459.
%
%
\bibitem{K-N_PRD15} 
Y.~Koide and H.~Nishiura,  Phys.~Rev. {\bf D 91} (2015) 116002. 
%
\bibitem{K-N_17}
Y.~Koide and H.~Nishiura, in preparation. 
%
\end{thebibliography}
\end{document}